%% file: main.tex
\documentclass[letterpaper]{article} 
\usepackage{aaai24}  
\usepackage{times}  
\usepackage{helvet}  
\usepackage{courier}  
\usepackage[hyphens]{url}  
\usepackage{graphicx} 
\usepackage{natbib}  
\usepackage{caption} 
\usepackage{newfloat}
\usepackage{algorithm}
\usepackage{algorithmic}

\usepackage{listings}
\DeclareCaptionStyle{ruled}{labelfont=normalfont,labelsep=colon,strut=off} 
\floatstyle{ruled}
\newfloat{listing}{tb}{lst}{}
\floatname{listing}{Listing}
\usepackage{algorithm}
\usepackage{algorithmic}
\usepackage{tabularx}
\usepackage{subfigure}
\usepackage{multirow}
\usepackage{amsmath}
\usepackage{amsfonts}
\usepackage{amsthm}
\usepackage{makecell}
\usepackage{booktabs}
\usepackage{colortbl}
\usepackage{enumerate}
\usepackage{enumitem}
\usepackage{lipsum}
\usepackage{multibib}

\def\model{SKES}
\newcommand\mth{\small}
\def\emb{\boldsymbol} 
\DeclareMathOperator\argmin{argmin}

\definecolor{darkblue}{RGB}{0, 20, 255}
\definecolor{light}{RGB}{23, 107, 135}
\definecolor{cell}{RGB}{174, 226, 255}

\newenvironment{sequation}{\begin{equation}\small\setlength\abovedisplayskip{2pt}\setlength\belowdisplayskip{2pt}}{\end{equation}}

\title{Deep Structural Knowledge Exploitation and Synergy for Estimating Node Importance Value on Heterogeneous Information Networks}

\author {
    Yankai Chen\textsuperscript{\rm 1}, 
    Yixiang Fang\textsuperscript{\rm 2}\thanks{The corresponding author.}, 
    Qiongyan Wang\textsuperscript{\rm 3}, 
    Xin Cao\textsuperscript{\rm 4}, 
    Irwin King\textsuperscript{\rm 1}
}
\affiliations {
    \textsuperscript{\rm 1} The Chinese University of Hong Kong, Hong Kong \\
    \textsuperscript{\rm 2} The Chinese University of Hong Kong, ShenZhen\\
    \textsuperscript{\rm 3} University of Copenhagen, Denmark\\
    \textsuperscript{\rm 4} The University of New South Wales, Australia\\
    \{ykchen,king\}@cse.cuhk.edu.hk, fangyixiang@cse.cuhk.edu.cn, qiongyanwang010@gmail.com, xin.cao@unsw.edu.au
}

\begin{document}

\maketitle

\input{sections/abstract}

\input{sections/intro}

\input{sections/pre}

\input{sections/method/method}


\input{sections/exp/exp}

\input{sections/con}

\section{Acknowledgments}
The work described here was partially supported by grants from the Research Grants Council of the Hong Kong Special Administrative Region, China (CUHK 14222922, RGC GRF No. 2151185) and (RGC Research Impact Fund R5034-18; CUHK 2410021).
Yixiang Fang was supported in part by NSFC (Grant 62102341), Guangdong Talent Program (Grant 2021QN02X826), Shenzhen Science and Technology Program (Grants JCYJ20220530143602006 and ZDSYS 20211021111415025), and Shenzhen Science and Technology Program and Guangdong Key Lab of Mathematical Foundations for Artificial Intelligence.


\bibliography{ref}

\end{document}

%% file: sections/abstract.tex
\begin{abstract}
\textit{Node importance estimation} problem has been studied conventionally with homogeneous network topology analysis.
To deal with network heterogeneity, a few recent methods employ graph neural models to automatically learn diverse sources of information.
However, the major concern revolves around that their full adaptive learning process may lead to insufficient information exploration, thereby formulating the problem as the \textit{isolated} node value prediction with underperformance and less interpretability.
In this work, we propose a novel learning framework: \model.
Different from previous automatic learning designs, \model~exploits heterogeneous structural knowledge to enrich the informativeness of node representations.
Based on a \textit{sufficiently uninformative} reference, \model~estimates the importance value for any input node, by \textit{quantifying} its disparity against the reference.
This establishes an interpretable node importance computation paradigm.
Furthermore, \model~dives deep into the understanding that ``\textit{nodes with similar characteristics are prone to have similar importance values}'' whilst guaranteeing that such informativeness disparity between any different nodes is orderly reflected by the embedding distance of their associated latent features.
Extensive experiments on three widely-evaluated benchmarks demonstrate the performance superiority of \model~over several recent competing methods.
\end{abstract}

%% file: sections/intro.tex
\section*{Introduction}
 \label{s:intro}

Estimating node importance, as one of the classic problems in network science, founds various downstream applications, such as recommender systems, web information search and retrieval, query disambiguation, and resource allocation optimization~\cite{zhang2019doc2hash,park2020multiimport,zheng2021convolutional,yang2022hicf,zhang2022knowledge,hu2020selfore,hu2022chef,hu2021gradient,DBLP:conf/cikm/SongZK23,chen2021efficient,fang2017effective,he2023dynamically,he2023dynamic}.
Traditional approaches revolve around the analyses of \textit{network topologies}, e.g., closeness centrality~\cite{nieminen1974centrality}, degree analysis~\cite{nieminen1974centrality}, and PageRank methodologies~\cite{page1999pagerank,haveliwala2003topic}.

\begin{figure}[t]
 \centering
 \includegraphics[width = 1\linewidth]{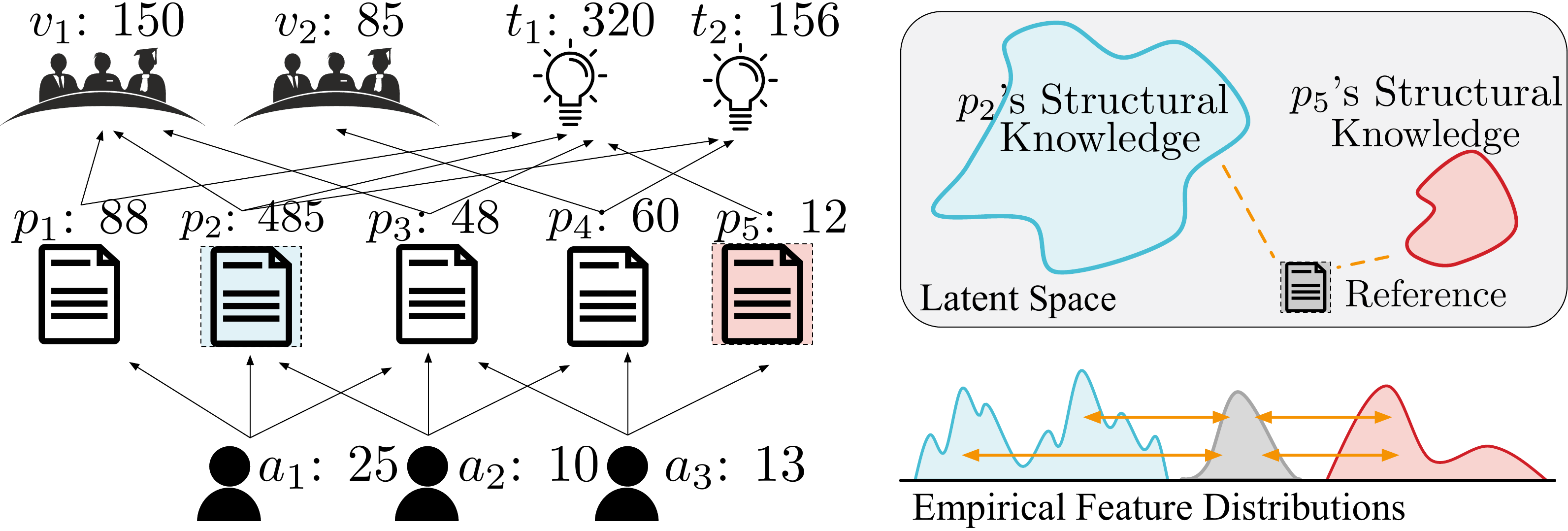}
 \caption{An HIN example and \model~methodology.} 
  \label{fig:intro}
\end{figure}

With the proliferation of heterogeneous information in graph data, conventional methods focusing on topology analysis may thus fail to capture such diverse semantic knowledge embedded. 
{Heterogeneous Information Networks} (HINs), which are essentially prevalent in various domains including bibliographic information networks, social media, and knowledge graphs, usually are composed of multiple typed nodes and edges.
Notably, \textbf{network heterogeneity leads to the variations of semantics and values in interpreting node importance}. 
This makes the studied problem more challenging on HINs than that on homogeneous counterparts.
We illustrate this by an HIN example of DBLP network in Figure~\ref{fig:intro}(a).
The important values of \textit{authors}, \textit{papers}, \textit{venues}, and \textit{topics} are indicated by their h-index values, citation numbers, venue rank, and popularity (e.g., numbers of Web pages in Google), respectively.
Despite the network connection, adjacent nodes may have different influences on the importance value of the target node. 
For example, \textit{authors} and \textit{topics} have different facets of contributions to the importance values of \textit{papers}.
Moreover, while the scholar's h-index values are often in [0, 350]\footnote{\scriptsize https://www.webometrics.info/en/hlargerthan100}, the maximum paper citation number could be over 300,000\footnote{\scriptsize 
 https://www.genscript.com/top-100-most-cited-publications.html}.
This shows that the importance value heterogeneity is essentially influenced by semantic heterogeneity. 
Consequently, it becomes evident that both variations are inherently difficult to be simply analyzed from homogeneous network topologies, necessitating the exigency for new methodologies.

\textbf{Related Works.}
Among a few recent attempts, methods with \textit{Graph} \textit{Neural} \textit{Networks} (GNNs) have emerged as a promising direction~\cite{park2019estimating}.
Due to good knowledge mining ability from high-order topologies, GNNs can produce semantic enrichment to vectorized node representations benefiting downstream tasks~\cite{bgr,lkgr,bgch,yang2023kappahgcn,yang2023hyperbolic,zhang2023mitigating,chen2020literature}. 
By incorporating specific designs for HINs, GNN-based methods show the potential in dealing with information heterogeneity~\cite{liang2023predicting,fu2023fedhgn}, especially for node importance estimation.
For instance, GENI~\cite{park2019estimating} applies GNN and attention mechanism to aggregate the structure information for node importance estimation.
MULTIIMPORT~\cite{park2020multiimport} improves GENI by using a variety of external input signals.
RGTN~\cite{huang2021representation} utilizes both the network structure information and nodes' attributes for estimating the importance of nodes.
However, all these works focus on solving the \textit{importance-based ranking problem}, without inferring the specific importance values.
Recent work HIVEN~\cite{ch2022hiven} considers the value heterogeneity of node importance in HINs by learning both local and global node information.
Nevertheless, the primary concern lies in that HIVEN purely relies on GNNs for automatic information aggregation but ignores the explicit structural knowledge mining on HINs, making the model underperforming and less interpretable in importance calculation.

\textbf{Our Contribution.}
We push forward the investigation of node importance estimation over HINs by introducing a novel learning framework, namely \model~(Deep \underline{S}tructural \underline{K}nowledge \underline{E}xploitation and \underline{S}ynergy).
\model~makes the assumption that each node corresponds to a unique \textit{high-dimensional feature distribution} reflecting its essential characteristics and knowledge to determine the node importance. 
However, such feature distribution is unknown and agnostic that can only be sampled and observed by certain \textit{empirical feature representations}. 
These empirical representations are expected to be as much informative with heterogeneous knowledge as possible, so that the importance of each node within the underlying HIN can thus be estimated from its associated feature representations. 
Then \model~transforms the importance regression problem into quantifying the \textit{informativeness} of these empirical node feature representations.
Underpinned by Optimal Transport Theory~\cite{villani2009optimal}, our formulation eventually provides an effectual and interpretable learning paradigm with theoretical guarantees. 

Specifically, \model~consists of three progressive modules. 
For each node, (1) \textit{Structural Priori Knowledge Exploitation} focuses on mining the intrinsic intra- and inter-node information, i.e., \textit{centrality} and \textit{similarity}, from the given HIN, providing the comprehensive coverage of structural knowledge with diversity and heterogeneity.
Then our (2) \textit{Synergetic Representation of Feature Distribution} module learns to empirically represent the node's unique, complicated, and high-dimensional feature distribution with adaptive heterogeneous knowledge synergy.
Lastly, (3) we manually create a random feature distribution as the reference, which functions as the ``coordinate origin'' in the embedding space.
Due to the randomness, this reference is \textit{sufficiently uninformative}.  
Then our \textit{Node Importance Value Estimation} module quantifies the informativeness of the input node by measuring its distance against the reference in the latent sapce, and transforms such measurement for node importance estimation.
Furthermore, anchoring on this reference, the estimated importance values obey \textit{triangle inequality} such that the informativeness gap between different node pairs can also be captured.
This produces a fine-grained importance learning framework, which is different from previous methods that formulate the problem as the importance value prediction for isolated nodes.
We provide a high-level illustration in Figure~\ref{fig:intro}(b) and summarize our principal contributions as:
\begin{itemize}[leftmargin=*]
\item To the best of our knowledge, we are the first to formulate the HIN node importance estimation problem via quantifying node feature informativeness with Optimal Transport methodology, providing a novel and interpretable perspective to the related community.

\item We propose \model~model with three effective modules that operate progressively from structural knowledge exploitation and synergy to node importance estimation.

\item We conduct model evaluation on three real-world benchmarks. Experimental results demonstrate the performance superiority of our model against competing methods as well as the effectiveness of each module contained therein.
\end{itemize}

%% file: sections/pre.tex
\begin{figure*}[h]
      \centering
    \includegraphics[width = 1\linewidth]{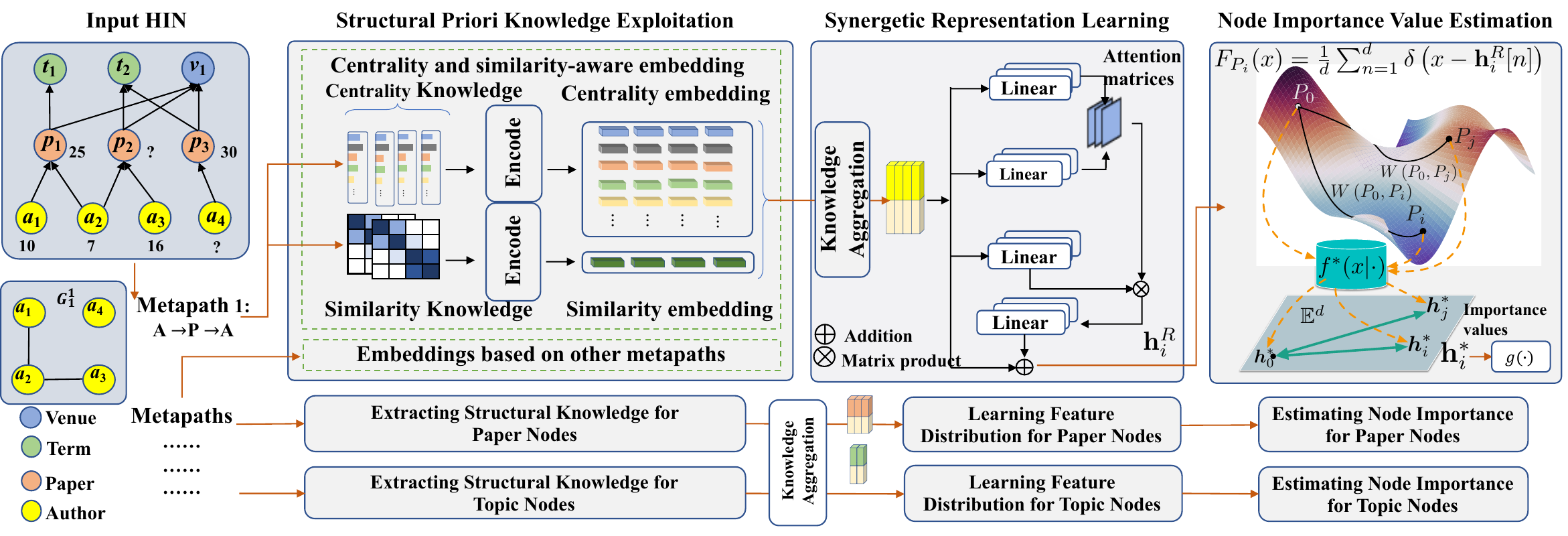}
    \caption{The framework of our proposed model (best view in color).} 
    \label{fig:Framework}
\end{figure*}

\section{Preliminaries and Problem Formulation}
\label{sec:pre}
\noindent\textbf{Definition\,1:}\,\textbf{Heterogeneous\,Information\,Network\,(HIN).} 
It is a directed graph $\mathcal H$ = ($\mathcal V$, $\mathcal E$) with a node type mapping function $\mathcal{\phi:V \xrightarrow{} A}$ and an edge type mapping function $\mathcal{\psi:E \xrightarrow{} R}$, where $\mathcal{A}$ is a set of node types and $\mathcal{R}$ is a set of edge types satisfying $|\mathcal{A}| + |\mathcal{R}| > 2$.

\noindent\textbf{Definition\,2: Metapaths.} 
A metapath $\mathcal P$ is with the form
{\small $A_1\stackrel{R_1}{\longrightarrow}A_2\stackrel{R_2}{\longrightarrow}\cdots\stackrel{R_H}{\longrightarrow}A_{H+1}$} that defines on the node and edge types, i.e., $A_i\in\mathcal A$, $R_i\in\mathcal R$.
We omit the edge types if they are unique between two connected node types, e.g., $A_1A_2\cdots A_{H+1}$.
We call a path between nodes $v_1$ to $v_{H+1}$ a {\it path instance} of $\mathcal P$, if $\forall i$, the node $v_i$ and edge $e_i$ = ($v_i$, $v_{i+1}$) satisfy that $\phi(v_i)$ = $A_i$ and $\psi(e_i)$ = $R_i$.

\noindent\textbf{Definition\,3: 1-Wasserstein Distance.} 
While Optimal Transport (OT) is the problem of moving one distribution of mass, e.g., {${P}$}, to another, e.g., {${Q}$}, as efficiently as possible, 1-Wasserstein distance is the derived minimum distribution distance that is defined by the following formulation:
\begin{sequation}
\label{eq:distance}
W\left(P, Q\right) = \inf _{f \in TP\left(P, Q\right)} \int \|\emb{x}-f(\emb{x})\|_1 d P(\emb{x}),
\end{sequation}%
where the infimum is over {\mth $TP(P, Q)$} that denotes all transport plans.
If a minimizer {\mth ${f}^*$} exists, it is thus the solution to compute $W(P, Q)$.
For common one-dimensional distributions, there is a closed-form solution to compute the optimal plan $f^*$ as $f^*(x):=F_P^{-1}\left(F_Q(x)\right)$; $F$ is the cumulative distribution function (CDF) associated with the underlying distribution.
1-Wasserstein distance satisfies \textit{positive-definiteness}, \textit{symmetry}, and \textit{triangle inequality}~\cite{nietert2022statistical,DBLP:conf/iclr/KorotinSB23,chen2023wsfe,chen2023topological,pswe}.

\noindent{\bf \underline{Problem Definition}.}
{\it Given an HIN $\mathcal{H}$
=($\mathcal{V}$, $\mathcal{E}$) and the importance values for a subset of nodes $\mathcal{V'} \subset \mathcal{V}$ of some given types  $\mathcal{A'} \subseteq \mathcal{A}$, we aim to learn a mapping function $g(\cdot): \mathcal{V} \rightarrow{} \mathbb{R}$ that estimates the importance value of every node of the given types $\mathcal{A}'$ in $\mathcal{H}$.}

%% file: sections/method/method.tex
\section{\model~Methodology}
\label{sec:model}

\subsection{Overview}
We now formally introduce our \model~model (Deep \underline{S}tructural \underline{K}nowledge \underline{E}xploitation and \underline{S}ynergy).
To implement the mapping function $g(\cdot)$, the general framework of \model~to estimate the node importance value is: 
\textit{given a random distribution $P_0$ representing the sufficiently uninformative reference, for each node $v_i$ $\in$ $\mathcal{V}$, we first learn to represent the high-dimensional feature distribution $P_i$ for $v_i$, then the mapping function $g(\cdot)$ can be implemented as $\forall v_i$ $\in$ $\mathcal{V}, $ $g(v_i)$ $=$ $g(W(P_0, P_i))$}.
As depicted in Figure~\ref{fig:Framework}, \model~comprises three progressively-operated modules.

\input{sections/method/knowledge_extraction}

\input{sections/method/synergy}

\input{sections/method/estimation}

\input{sections/method/training}

%% file: sections/method/knowledge_extraction.tex
\subsection{Structural Priori Knowledge Exploitation}
Acquiring informative structural knowledge is critical to estimating node importance.
To achieve this in HINs, we adopt the metapath-based methodologies~\cite{sun2011pathsim} to first obtain the underlying sub-networks as follows.

\subsubsection{\emph{1) Metapath-induced Sub-network Construction.}}
There are limited but meaningful \textit{metapaths} in HINs that describe the meta information of HINs~\cite{fu2020magnn,fu2024mecch}.
For the $k$-th metapath, the induced sub-network is denoted as $G^{j}_k$ = ($\mathcal V_k^{j}$, $\mathcal E^{j}_k)$,
such that $\mathcal V_k^j$ contains all $j$-th typed nodes and $\mathcal E_k^j$ contains all the edges between nodes in $\mathcal V_k^j$, i.e., two nodes are linked in this sub-network if there is a path instance of the metapath between them.

\subsubsection{\emph{2) Priori Centrality and Similarity Embedding.}}
After obtaining the induced sub-networks, we propose to extract node pointwise \textit{centrality} and pairwise \textit{similarity}, which are two essential network properties that reveal the \textit{intra-} and \textit{inter-}node priori knowledge.
Specifically, centrality measures generally are either defined based on network properties~\cite{nieminen1974centrality,hirsch2005index,egghe2006improvement,negre2018eigenvector,page1999pagerank,dorogovtsev2006k}, or designed based on the shortest paths~\cite{shaw1954group,marchiori2000harmony,sabidussi1966centrality}.
To fully capture the diverse information of the underlying structures, we pre-process these popular centrality measures to improve knowledge coverage.
Detailed formulations are listed in Appendix.
Then each centrality value is vectorized into 128-dimension via a two-layer perceptron.
We denote the embedding calculated from the $l$-th ($l$ ranges from 1 to $L$) centrality value of node $v_i$ by {\small $\mathbf c_{i,k}^{(l)}$}.

While centrality reflects the property of a given node oneself, \textit{similarity} reveals the contrastive node information compared to others.
In this work, \model~embeds the knowledge from the \textit{attribute-} and \textit{topology-aware} similarity. 
Specifically, for each edge in the induced sub-network $G^{j}_k$, we assign the edge weight by the cosine similarity between the attributes of its two end nodes. Then we transform the similarity matrix as the transition probabilities to compute the node embeddings by adopting node2vec~\cite{grover2016node2vec}.
To embed topology-aware similarity, we directly follow PathSim~\cite{sun2011pathsim} to calculate the similarity of each node pair in $G^{j}_k$ and then take an analogous embedding procedure via node2vec.
Subsequently, we take the summation of two corresponding embeddings, denoted by $\mathbf f_{i,k}^{att}$ and $\mathbf f_{i,k}^{top}$ of the node $v_i$, and finally have the similarity knowledge embedding as $\mathbf c_{i,k}^{(L+1)}=\mathbf f_{i,k}^{att}+\mathbf f_{i,k}^{top}$.

To summarize, centrality and similarity knowledge provides cohesive and complementary views of underlying networks, as each one of these measures usually represents a specific perspective of structural information.
As we will show in experiments, they substantially boost \model~performance and stabilize the model training.

%% file: sections/method/synergy.tex
\subsection{Synergetic Representation of Feature Distribution}

\subsubsection{\emph{1) Heterogeneous Knowledge Aggregation.}}
Since the priori centrality and similarity knowledge represents the node information from different perspectives, we propose to synergetically fuse them for later representing the unknown high-dimensional feature distribution.
Concretely, for each node $v_i$, let {\small $\alpha_{i,k}^{(l)}$} denote the weighting coefficient of the $l$-th centrality embedding and {\small $\alpha_{i,k}^{(L+1)}$} represent the coefficient of $v_i$'s similarity embedding.
We first derive the following coefficient calculations with $l$ ranging from 1 to $L+1$:
\begin{sequation}
\alpha_{i,k}^{(l)} = \frac{1}{|G_k^{\phi(v_i)}|} \sum_{v_{i'} \in G_k^{\phi(v_i)}}  \mathbf W_1  \tanh(\mathbf W_1' \mathbf c_{i^\prime,k}^{(l)}+ \mathbf b_1), \\
\end{sequation}%
where $G_k^{\phi(v_i)}$ denotes the induced sub-network.
$\mathbf W_1$, $\mathbf W_1'$, and $\mathbf b_1$ are learnable parameters. 
These coefficients are further normalized with the softmax function, i.e., {\small $ \widehat{\alpha}_{i,k}^{(l)}$ $=$ $\exp(\alpha_{i,k}^{(l)})/ \sum_{l^\prime=1}^{L+1} \exp(\alpha_{i,k}^{(l^\prime)})$}.
{\small $\widehat{\alpha}_{i,k}^{(l)}$} attentively contributes to the $k$-th metapath derived knowledge embedding $\mathbf e_{i,k}$ as:
\begin{small}
\begin{sequation}
\mathbf e_{i,k} = \sum_{l=1}^{L+1} \widehat{\alpha}_{i,k}^{(l)} \mathbf c_{i,k}^{(l)}.
\end{sequation}%
\end{small}

Besides, for each node $v_i$, since different metapaths lead to different sub-network extraction, we further adaptively fuse embeddings from these different sub-networks containing $v_i$ as well.
Similarly, the coefficient $\tau_{i,k}$ of $v_i$ induced by the $k$-th metapath is defined as:
\begin{sequation}
\label{eq:eq1}
 \tau_{i,k} = \frac{1}{|G_k^{\phi(v_i)}|} \sum_{v_{i'} \in G_k^{\phi(v_i)}}  \mathbf W_2  \tanh(\mathbf W_2' \mathbf e_{i^\prime,k}+ \mathbf b_2),
\end{sequation}%
where $\mathbf W_2$, $\mathbf W_2'$, and $\mathbf b_2$ are learnable parameters. 
Similarly, Eqn.~(\ref{eq:eq1}) is further normalized across all other related coefficients as, {\small $\widehat{\tau}_{i,k}$ = $\exp(\tau_{i,k})/\sum_{k^\prime=1}^{N_{\phi(v_i)}} \exp(\tau_{i,k^\prime})$}.
{\small $N_{\phi(v_i)}$} denotes the number of metapaths starting from node type $\phi(v_i)$.
We derive the aggregated embedding of $\mathbf e_{i}$ as follows:
\begin{sequation}
\mathbf e_{i} =\sum_{k=1}^{N_{\phi(v_i)}} \widehat{\tau}_{i,k} \mathbf e_{i,k}.
\end{sequation}%
Given $v_i$'s initial feature $\mathbf e'_i$ and the learned knowledge embedding $\mathbf e_i$, we obtain the aggregated representation $\mathbf x_i$ by concatenation (denoted as $||$), i.e., $\mathbf x_i$ = $\mathbf e'_i$ $||$ $\mathbf e_i$.

\subsubsection{\emph{2) Empirical Representation of Feature Distribution.}}
Intuitively, $\mathbf x_i$ contains both the initial node features and refined structural knowledge.
We then adaptively learn the empirical representations that are informative to represent the \textit{unknown high-dimensional node feature distributions}.

We achieve this by leveraging the self-attention mechanism~\cite{vaswani2017attention,chen2022attentive}.
Specifically, we extract hidden features from the input $\textbf x_i$ via implementing $M$ attention heads in each layer.
We denote the hidden feature of node $v_i$ learned by the $r$-th layer as $\mathbf h_i^{(r)}$. 
We follow the conventional computation protocol to firstly obtain $d$-dimensional \textit{query}, \textit{key} and \textit{value} variables:
\begin{sequation}
\label{eq:trsfm}
\resizebox{1\linewidth}{!}{$
   \mathbf q_{i,m}^{(r)} = \mathbf W_{qry}^m \mathbf h_{i,m}^{(r)},\,
   \mathbf k_{i,m}^{(r)} = \mathbf W_{key}^m \mathbf h_{i,m}^{(r)},\,
   \mathbf v_{i,m}^{(r)} = \mathbf W_{val}^m \mathbf h_{i,m}^{(r)},
  $}
\end{sequation}%
where {\small $\mathbf q_{i,m}^{(r)}$}, {\small$\mathbf k_{i,m}^{(r)}$}, and {\small$\mathbf v_{i,m}^{(r)}$} denote the $m$-th query, key, and value vectors of $v_i$ at the $r$-th layer.
$\mathbf W_{qry}^m$, $\mathbf W_{key}^m$, and $\mathbf W_{val}^m$ are learnable weights. 
Then, the attentive coefficient between nodes $v_j$ and $v_i$ is calculated as:
\begin{sequation}
\label{eq:trsfm_att}
\resizebox{1\linewidth}{!}{$
S_m^{(r)}(v_j, v_i) = \frac{\exp(\mathbf q_{i,m}^{(r)} \mathbf W_{\psi(e_{j,i})}(\mathbf k_{i,m}^{(r)})^{\mathsf{T}} \frac{\mu^{\psi(e_{j,i})}} {\sqrt{d}} )  } {  \underset{v_{j^\prime} \in \mathcal N(v_i)}{\sum} \exp(\mathbf q_{i,m}^{(r)} \mathbf W_{\psi(e_{j^\prime,i})}(\mathbf k_{j^\prime,m}^{(r)})^{\mathsf{T}} \frac{\mu^{\psi(e_{j^\prime,i})} } {\sqrt{d}} ) },
$}
\end{sequation}%
where $e_{j,i}$ denotes the edge from $v_j$ to $v_i$, and {\small$\mathbf W_{\psi(e_{j,i})}$} represents the learnable weight matrix of edge type {\small$\psi(e_{j,i})$}.
{\small$\mu^{\psi(e_{j,i})}$} is the learnable magnitude for type {\small$\psi(e_{j,i})$} and {\small$\mathcal N(i)$} is the neighbor set of node $v_i$.
Then the embedding {\small$\mathbf v_{i,m}^{(r)}$} is updated via aggregating adjacent information as follows: 
\begin{sequation}
\tilde{\mathbf v}_{i,m}^{(r)} = \underset{v_j \in \mathcal N(v_i)}{\sum} S^{(r)}_m(v_j,v_i) \mathbf v_{i,m}^{(r)}. 
\end{sequation}%
Let $||$ denote the concatenation and $\mathbf W_{out}$ is a learnable weight matrix.
We finally complete the target feature representation by iteratively updating from $r=1$ to $R-1$:
\begin{sequation}
\label{eq:trsfm_iter}
{\mathbf h}_{i,m}^{(r+1)} = \mathbf h_{i,m}^{(r)} + \mathbf W_{out} \cdot \left({||}_{m=1}^M \tilde{\mathbf v}_{i,m}^{(r)}\right).
\end{sequation}%
The output of Eqn.~(\ref{eq:trsfm_iter}), i.e., ${\mathbf h}_i^{R}$ for brevity, is expected to be empirically representative for the \textit{unknown node feature distribution} with heterogeneous knowledge synergy.
We explain our implementation to estimate node importance via mensurating the empirical distribution distances as follows.

%% file: sections/method/estimation.tex
\subsection{Node Importance Value Estimation}
As we mentioned earlier, notation $P_i$ denotes the feature distribution associated with node $v_i$.
Since we use ${\mathbf h}_i^R$ to represent $P_i$, which is discrete, then its empirical CDF can be defined as follows:
\begin{sequation}
\label{eq:ecdf}
F_{{P}_i}(x)=\frac{1}{d} \sum_{n=1}^{d} \delta\left(x-{\mathbf h}_i^{R}[n]\right),
\end{sequation}%
where {\mth $\delta(\cdot)$} returns 1 if the input is zero and 0 otherwise\footnote{Dirac delta function with $\int$$\delta(x)dx$ $=$ $1$ for continuous inputs.}.
${\mathbf h}_i^{R}[n]$ is the $n$-th element.
To explicitly measure the distribution distance, we propose to compare $P_i$ with a fixed \textit{random reference} that functions as the ``origin'' in the embedding space.
Specifically, we introduce a reference distribution $P_{0}$ with associated feature representation ${\mathbf h}_0$ $\in$ $\mathbb{R}^d$, elements of which are uniformly sampled. 
Then the distribution distance between $P_i$ and $P_0$ can be explicitly measured by 1-Wasserstein distance, i.e., $W(P_0, P_i)$.

\input{sections/tables/dataset}

As we have introduced in Preliminaries, $W(P_0, P_i)$ is computed via implementing the optimal transport plan $f^*(x):=F_{P_i}^{-1}\left(F_{P_0}(x)\right)$. Based on the empirical CDFs of $P_0$ and $P_i$, we can quantitatively interpreted $f^*(x)$ as:
\begin{sequation}
f^*\left(x | {\mathbf h}_i^{R}\right)={\argmin}_{x^{\prime} \in {\mathbf h}_i^{R}}\left(F_{P_i}(x^{\prime})=\gamma\right), \text{ } \gamma=F_{P_0}(x).
\end{sequation}%
Moreover, let $\pi(x^{\prime} | {\mathbf h}_i^{R})$ denote the ranking of each input $x^{\prime}$ in the ascending sorting of elements in ${\mathbf h}_i^{R}$. We can further achieve the following algorithmic implementation:
\begin{sequation}
f^*\left(x | {\mathbf h}_i^{R}\right)={\argmin}_{x^{\prime} \in {\mathbf h}_i^{R}}\left(\pi(x^{\prime} | {\mathbf h}_i^{R}) = \pi(x | {\mathbf h}_0)\right).
\end{sequation}%
Please notice that, the indicator $\pi(\cdot)$ can be actually pre-processed via ``argsort'' to ${\mathbf h}_i^{R}$ and ``sort'' to ${\mathbf h}_0$, which \textit{essentially permutes and encodes ${\mathbf h}_i^{R}$ via referring to ${\mathbf h}_0$}.
Therefore, the resultant representation is denoted as ${\mathbf h}_i^{*}$:
\begin{sequation}
{\mathbf h}_i^{*} = ||_{n=1}^d \left( f^*({\mathbf h}_0[n] | {\mathbf h}_i^{R})-{\mathbf h}_0[n] \right).
\end{sequation}%
${\mathbf h}_i^{*}$ $\in$ $\mathbb{R}^d$ presents several desirable geometric properties, as it can efficiently reflect the 1-Wasserstein distance between distributions $P_0$ and $P_i$ as follows:
\begin{sequation}
\label{eq:w0i}
\left\| {\mathbf h}_i^{*} \right\|_1 \propto W(P_0, P_i)  \text{ and } \left\| {\mathbf h}_i^{*} - {\mathbf h}_j^{*} \right\|_1  \propto  W(P_i, P_j).
\end{sequation}%

We attach the proof of Eqn.~\ref{eq:w0i} in Appendix.
${\mathbf h}_i^{*}$ naturally inherit the relative order of the distance ranking with the theoretical guarantees. 
Based on ${\mathbf h}_i^{*}$, we finally implement the importance estimation function $g(\cdot)$:
\begin{sequation}
\label{eq:score}
g(v_i) =  \emb{\lambda} \cdot {\mathbf h}_i^{*},
\end{sequation}%
where $\emb{\lambda}$ is a learnable vector to provide better regression capability.  
Obviously, for any node $v_i$, its importance value $g(v_i)$ is correlated with its distribution distance to the reference $P_0$.
We showcase and analyze its performance superiority over competing methods in Experimental Evaluation.

%% file: sections/tables/dataset.tex
\begin{table*}[t]
\fontsize{9}{9} \selectfont
\centering
\setlength{\tabcolsep}{1.3mm}{
    \begin{tabular}{c|c|c|c|c|c|c|c|c}
    \specialrule{.1em}{0em}{0em}
    Dataset  & \# Nodes & \# Edges  & \# Node types & \# Edge types & Target node & Meaning & \# Node with Importance & Range \\ \specialrule{.1em}{0em}{0em}
    \multirow{2}{*}{MUSIC10K} & \multirow{2}{*}{22,986}  & \multirow{2}{*}{80,272}  & \multirow{2}{*}{4} & \multirow{2}{*}{8}  & Artist & Familiarity & 4,214 (18.3\%) & [0, 1] \\ \cline{6-9} 
             &         &         &   &    & Song     & Hotness     & 4,411 (19.1\%) & [0, 1] \\ \hline
    
    \multirow{2}{*}{TMDB5K}   & \multirow{2}{*}{76,926}  & \multirow{2}{*}{359,780} & \multirow{2}{*}{7} & \multirow{2}{*}{12} & Movie    & Popularity  & 4,802 (6.2\%)  & [-7.89, 6.77]  \\  \cline{6-9}
             &         &           &   &    & Director & Box office  & 1,159 (1.5\%)  & [0.021, 10.55]  \\ \hline
    \multirow{2}{*}{DBLP}     & \multirow{2}{*}{249,903} & \multirow{2}{*}{2,428,250} & \multirow{2}{*}{4} & \multirow{2}{*}{6}  & Author   & H-index & 101,958 (40.8\%)  & [0, 159]   \\ \cline{6-9}
             &         &           &   &    & Paper    & Citations & 100,000 (40.0\%)  & [0, 34191] \\  \specialrule{.1em}{0em}{0em}
    \end{tabular}
}
\caption{ Statistics of three datasets (MUSIC10K, TMDB5K and DBLP).}
\label{tab:data}
\end{table*}

%% file: sections/method/training.tex
\subsection{Training Objective}
\label{sec:train}

We adopt the common regression loss with mean squared error between estimated and ground-truth importance values:
\begin{sequation}
\label{eq:mse}
 \mathcal{L}_{mse}=\frac{1}{|\mathcal A^{\prime}|}\sum_{j\in \mathcal A^{\prime} } \frac{1}{|{\mathcal V}^j|}\sum_{v_i\in {\mathcal V}^j} (g(v_i)-y_{v_i})^{2},
\end{sequation}%
where $y_{v_i}$ denotes the ground-truth importance value of node $v_i$.
Then the complete training objective is defined as:
\begin{sequation}
 \mathcal{L} = \mathcal{L}_{mse} + \mu\|\Delta\|_2^2.
\end{sequation}%
$\|\Delta\|_2^2$ is the $L2$-regularizer of trainable embeddings and variables to avoid over-fitting with the hyperparameter $\mu$. 

%% file: sections/exp/exp.tex
\input{sections/tables/importance}

\input{sections/tables/rank}

\section{Experiments}
\label{sec:exp}
We evaluate \model~with the research questions (RQs) as:
\begin{itemize}[leftmargin=*]
\item \textbf{RQ1}: How does \model~compare to state-of-the-art methods on the tasks of \textit{node importance value estimation} and \textit{important node ranking}?

\item \textbf{RQ2}: How does our proposed design of knowledge synergy and measurement contribute to \model~performance?

\item \textbf{RQ3}: How does other proposed components of \model~influence the model performance?
\end{itemize}

\input{sections/exp/setups}

\input{sections/exp/model_performance}

\input{sections/exp/study}

\input{sections/exp/ablation}

%% file: sections/tables/importance.tex
\begin{table*}[t]
\centering
\fontsize{9.5}{9.5} \selectfont
\setlength{\tabcolsep}{0.1mm}{
 \begin{tabular}{c|c c c|c c c|c c c|c c c|c c c|c c c}
\specialrule{.1em}{0em}{0em}
\multirow{3}{*}{Method}&\multicolumn{6}{c|}{MUSIC10K}&\multicolumn{6}{c|}{TMDB5K}&\multicolumn{6}{c}{DBLP}\\ \cline{2-7} \cline{8-13} \cline{14-19}
&\multicolumn{3}{c|}{Song}&\multicolumn{3}{c|}{Artist}&\multicolumn{3}{c|}{Movie}&\multicolumn{3}{c|}{Director}&\multicolumn{3}{c|}{Paper}&\multicolumn{3}{c}{Author}\\ \cline{2-4} \cline{5-7} \cline{8-10} \cline{11-13} \cline{14-16} \cline{17-19} 
&M&R&N&M&R&N&M&R&N&M&R&N&M&R&N&M&R&N \\
\specialrule{.1em}{0em}{0em}
PR&0.460&0.489&0.633&0.540&0.563&0.596&2.517&2.771&0.448&0.499&1.005&0.155&2.293&3.244&0.352&1.728&1.894&0.467\\
PPR&0.460&0.490&0.633&0.540&0.563&0.596&2.512&2.771&0.448&0.499&1.005&0.155&2.293&  3.243&0.352&1.173&1.894&0.467\\
\hline
LR&0.137&0.165&0.208&0.126&0.164&0.173&0.672&0.851&0.138&0.541&0.812&0.125&1.061&1.319&0.146&0.628&0.7657&0.204\\
RF&0.122&\underline{0.152}&0.213  &0.110&0.143&0.151&0.774&0.954&0.154&0.500&0.857&0.132&1.057&1.312&0.145&0.555&0.674&0.180\\
NN&0.127&0.155&0.200&0.111&0.143&0.151&0.720&0.889&0.144&0.402&0.801&0.124&1.042&1.297&0.141&1.133&1.402&0.153\\
GAT&0.126&0.156&0.205&0.109&0.140&0.149&0.635&0.808&0.131&0.415&0.753&0.116&0.991&\underline{1.240}&0.143&\underline{0.492}&\underline{0.606}&0.153\\
\hline
HGT&0.129&0.160&0.207&0.118&0.145&0.154&0.581&0.764&0.126&0.352&0.681&0.105&0.996&1.248&\underline{0.135}&0.514&0.629&0.158\\
GENI&0.131&0.158&\underline{0.200}    &0.121&0.155&0.158&0.594&0.748&0.121&0.347&0.677&0.104&1.006&1.259&0.145&0.493&0.607&0.154\\
MULTI&0.147&0.185&0.234&0.147&0.186&0.190&0.982&1.166&0.189&0.479&0.764&0.117&
2.061&2.451&0.268&1.316&1.467&0.371\\
RGTN&0.123&0.155&0.216&0.111&0.143&0.153&0.624&0.798&0.182&0.316&0.557&0.093&\underline{0.990}&1.240&0.136&0.496&0.613&0.155\\
HIVEN&\underline{0.122}&0.152 &0.209  &\underline{0.102}&\underline{0.132}&\underline{0.144}&\underline{0.523}&\textbf{0.664}&\underline{0.108}&\textbf{0.268}&\underline{0.539}&\underline{0.084}&1.024&1.283&0.141&0.507&0.620&\underline{0.151}\\
\hline
\textbf{\model} &\textbf{0.106}&\textbf{0.137}&\textbf{0.177}&\textbf{0.099}&\textbf{0.126}&\textbf{0.133}&\textbf{0.509}& \underline{0.667}&\textbf{0.106}& \underline{0.274}&\textbf{0.507}&\textbf{0.083}&\textbf{0.940}&\textbf{ 1.179}&\textbf{0.130}&\textbf{0.462}&\textbf{0.571}&\textbf{0.143}\\

Gain&{15.09\%}&{10.94\%}&{12.99\%}&{3.03\%}&{4.76\%}&{8.27\%}&{2.75\%}&{-0.45\%}&{1.89\%}&{-2.19\%}&{6.31\%}&{1.20\%}&{5.32\%}&{5.17\%}&{3.85\%}&{6.49\%}&{6.13\%}&{5.59\%}\\
\specialrule{.1em}{0em}{0em}
\end{tabular}}
\caption{Quantitative comparison on the importance value estimation task. Bold and underlined digits are the best and second-best metric values (M, R, and N denote MAE, RMSE, and NRMSE, respectively).
}
\label{tab:comp-import}
\end{table*}

%% file: sections/tables/rank.tex
\begin{table*}[t]
\centering
\fontsize{9.5}{9.5} \selectfont
\setlength{\tabcolsep}{0.85mm}{
\begin{tabular}{c|p{1.3cm}<{\centering} p{1.1cm}<{\centering}|p{1.3cm}<{\centering} p{1.1cm}<{\centering}|p{1.3cm}<{\centering} p{1.1cm}<{\centering}|p{1.3cm}<{\centering} p{1.1cm}<{\centering}|p{1.3cm}<{\centering} p{1.1cm}<{\centering}|p{1.3cm}<{\centering} p{1.1cm}<{\centering}}
    \specialrule{.1em}{0em}{0em}
    \multirow{3}{*}{Method}&\multicolumn{4}{c|}{MUSIC10K}&\multicolumn{4}{c|}{TMDB5K}&\multicolumn{4}{c}{DBLP}\\ \cline{2-5} \cline{6-9} \cline{10-13} 
    &\multicolumn{2}{c|}{Song}&\multicolumn{2}{c|}{Artist}&\multicolumn{2}{c|}{Movie}&\multicolumn{2}{c|}{Director}&\multicolumn{2}{c|}{Paper}&\multicolumn{2}{c}{Author}\\ \cline{2-3} \cline{4-5} \cline{6-7} \cline{8-9} \cline{10-11} \cline{12-13}     
    &SP&NDCG&SP&NDCG&SP&NDCG&SP&NDCG&SP&NDCG&SP&NDCG\\
    \specialrule{.1em}{0em}{0em}
    PR&0.013&0.596&0.176&0.743&0.548&0.775&0.182&0.473&-0.104&0.331&0.443&0.916\\
    PPR&-0.020&0.581&0.188&0.732&0.707&0.846&0.195&0.489&0.051&0.333&0.453&0.913\\
    \hline
    LR&0.226&0.701&-0.037&0.645&0.669&0.858&0.393&0.672&0.312&0.538&0.2445&0.676\\
    RF&0.461&0.797&0.441&0.783&0.590&0.854&0.333&0.484&0.325&0.615&0.396&0.759\\
    NN&0.383&0.774&0.431&0.820&0.657&0.850&0.414&0.613&0.352&0.583&-0.002&0.399\\
    GAT&0.408&0.786&0.481&0.830&0.728&0.867&0.660&0.794&0.401&0.597&0.491&0.922\\
    \hline
    HGT&0.342&0.753&0.448&0.810&0.758&0.892&0.301&0.463&{0.426}&\underline{0.644}&0.458&0.857\\
    GENI&0.402&0.793&0.485&0.784&0.753&0.895&0.678&0.851&0.412&0.602&\underline{0.491}&\underline{0.923}\\
    MULTI&0.467&0.808&0.500&0.871&0.728&0.867&0.660&0.704&0.364&0.596&0.452&0.918\\
    RGTN&0.414&0.787&0.486&0.853&0.682&0.901&0.623&0.822&\underline{0.438}&0.643&0.488&0.907\\
    HIVEN&\underline{0.480}&\underline{0.814}&\underline{0.544}&\underline{0.885}&\underline{0.793}&\underline{0.910}&\textbf{0.701}&\textbf{0.862}&0.404&0.612&0.459&0.913\\
    \hline
    \textbf{\model} &\textbf{0.565}&\textbf{0.865}&\textbf{0.602}&\textbf{0.894}&\textbf{0.823}&\textbf{0.942}&\underline{0.680}&\underline{0.847}&\textbf{0.483}&\textbf{0.674}&\textbf{0.589}&\textbf{0.925}\\
    Gain&{17.77\%}&{6.27\%}&{10.66\%}&{1.02\%}&{3.78\%}&{3.52\%}&{-3.00\%}&{-1.74\%}&{10.27\%}&{4.66\%}&{19.96\%}&{0.22\%}\\

    \specialrule{.1em}{0em}{0em}
\end{tabular}
}
\caption{ Quantitative comparison on the importance-based node ranking task (SP denotes SPEARMAN).
}
\label{tab:comp-rank}
\end{table*}

%% file: sections/exp/setups.tex
\subsection{Experiment Setups}

\subsubsection{\emph{1) Benchmarks.}}
We include three widely evaluated real-world HIN datasets, namely MUSIC10K, TMDB5K, and DBLP.
Dataset statistics are reported in Table~\ref{tab:data} and detailed data descriptions are attached in Appendix.

\subsubsection{\emph{2) Evaluation Metrics.}}
For the node importance estimation task, three metrics are applied for performance evaluation, including mean absolute error (MAE), root mean square error (RMSE), and normalized root mean square error (NRMSE). 
The lower value of these metrics, the better the model performance.
On the importance-based ranking task, we use normalized discounted cumulative gain (NDCG) and Spearman correlation coefficient (SPEARMAN), where a higher value indicates better performance.

\subsubsection{\emph{3) Experimental Settings.}}%
In line with prior work~\cite{park2019estimating,ch2022hiven}, we perform five-fold cross validation for testing and report the average performance.
For each fold, 80\% and 20\% of nodes with ground-truth importance values are used for training and testing. 15\% of training nodes is used for validating.
We select symmetric metapaths with lengths less than four.
We implement \model~using Python 3.8 and PyTorch 1.8.0 on a Linux machine with 4 Nvidia A100 GPUs and 4 Intel Core i7-8700 CPUs.
Following HIVEN~\cite{ch2022hiven}, node features are initialized from textual contents via sentence-BERT~\cite{reimers2019sentence}.
We set the learning rate as $10^{-3}$ and train the model via Adam optimizer.
Metapaths and hyperparameter settings are reported in Appendix.

\subsubsection{\emph{4) Competing Models.}}%
We include three groups of existing models:
(1) traditional network analytic methods, i.e., PageRank~(PR)~\cite{page1999pagerank} and personalized PageRank~(PPR)~\cite{haveliwala2003topic};
(2) machine learning methods, i.e., linear regression (LR) and random forest (RF);
(3) neural network based models, i.e., GAT~\cite{velivckovic2017graph}, HGT~\cite{hu2020heterogeneous}, GENI~\cite{park2019estimating}, Multiimport~(MULTI)~\cite{park2020multiimport}, RGTN~\cite{huang2021representation}, and HIVEN~\cite{ch2022hiven}.
Detail descriptions are referred in Appendix.

%% file: sections/exp/model_performance.tex
\subsection{Experimental Evaluation (RQ1)}
\label{sec:overall}
\textbf{\emph{1) Task of Importance Value Estimation.}}
As shown in Table \ref{tab:comp-import}, we observe that: 
\textbf{(1)} Compared to HGT, GENI, and MULTI that mainly learn the graph heterogeneity, HIVEN considers the degree centrality that showcases its usefulness for estimating the node importance.     
\textbf{(2)} \model~further outperforms the baseline methods with performance gains over 2.75\%, 4.76\%, and 1.20\% in MAE, RMSE, and NRMSE on three datasets. 
This demonstrates its effectiveness of structural knowledge exploitation and synergy for accurate estimation of importance values.
\textbf{(3)} For the target \textit{``Director''} type of TMDB5K, we observe a performance gap against HIVEN with -2.19\% of MAE.
One explanation is due to the {data scarcity} issue of \textit{``Director''} for model training, i.e., 1.5\% shown in Table~\ref{tab:data}, as \model~may be under-trained to produce sporadic inaccurate estimations.
These \textit{``outliers''} may significantly influence MAE that considers the average absolute difference between the predicted and actual values.
On the contrary, RMSE and NRMSE are less sensitive to these outliers, providing a more objective evaluation by considering the magnitude of errors and reflecting the influence of outliers less prominently.
\textbf{(4)} Additionally, we conduct the Wilcoxon signed-rank tests and the results show that all the improvements that \model~has achieved are statistically significant with at least 95\% confidence level.

\input{sections/tables/margin_ranking}

\textbf{\emph{2) Task of Important Node Ranking.}}
We present evaluation results for important node ranking in Table \ref{tab:comp-rank} with twofold discussions:
\textbf{(1)} Our model generally presents performance superiority over all baselines with 3.78\%$\sim$19.96\% and 0.22\%$\sim$6.27\% of metric improvements, respectively.
This is intuitive as our model \model, performing well for essential importance value estimation task, thus naturally inherits to present a good capability for ranking.
\textbf{(2)} 
For the type \textit{``Director''} of TMDB5K, \model~obtains the second-best performance.
Since our model incorporates more heterogeneous information, this implies that it may need more training data for better knowledge learning and fusion.

One solution could be to supplement the original regression objective, i.e., MSE loss of Eqn.~(\ref{eq:mse}), by adding \textit{triplet ranking objective}, e.g., marginal ranking loss, as follows:
\begin{sequation}
\label{eq:margin_loss}
\mathcal{L}_{mrl}=\max \big\{0, m + \big(g(v_i)-g(v_{+})\big) - \big(g(v_i)-g(v_{-})\big) \big\},
\end{sequation}%
where $m$ is the margin.
$v_{+}$ and $v_{-}$ are two nodes that the importance value gap between $v_{+}$ and $v_i$ is larger than the value gap between $v_{-}$ and $v_i$.
The general idea of Eqn.~(\ref{eq:margin_loss}) is to reduce the disparity between $v_i$ and its positive pair, i.e., $v_{+}$, against its negative counterpart $v_{-}$.
We conduct experiments on this case with results shown in Table~\ref{tab:margin_loss}.
The observation indicates that supplementing the ranking regularization of Eqn.~(\ref{eq:margin_loss}) will boost the performance as expected.
With the performance increasing over 6.49\%, our model also surpasses HIVEN shown in Table~\ref{tab:comp-rank}.
On the other hand, the training time cost inevitably increases, which is a practical trade-off to consider.
Hence, in this paper, we mainly report the results based on our original objective function and leave the further design of learning frameworks as future work.

%% file: sections/tables/margin_ranking.tex
\begin{table}[t]
\centering
\fontsize{9.5}{9.5} \selectfont
\setlength{\tabcolsep}{1.4mm}{
\begin{tabular}
{c|c|c}
\specialrule{.1em}{0em}{0em}
 SPEARMAN & NDCG & Training Time/epoch (ms) \\ \specialrule{.1em}{0em}{0em}
 0.680$\rightarrow$ 0.755 & 0.847$\rightarrow$0.902  & 745$\rightarrow$1,219   \\ 
(+11.03\%) & (+6.49\%) & (+63.62\%) \\
 \specialrule{.1em}{0em}{0em}
\end{tabular}
}
\caption{Evaluation with the marginal ranking loss.}
\label{tab:margin_loss}
\end{table}

%% file: sections/exp/study.tex
\begin{figure}[t]
    \centering
    \includegraphics[width=1\linewidth]{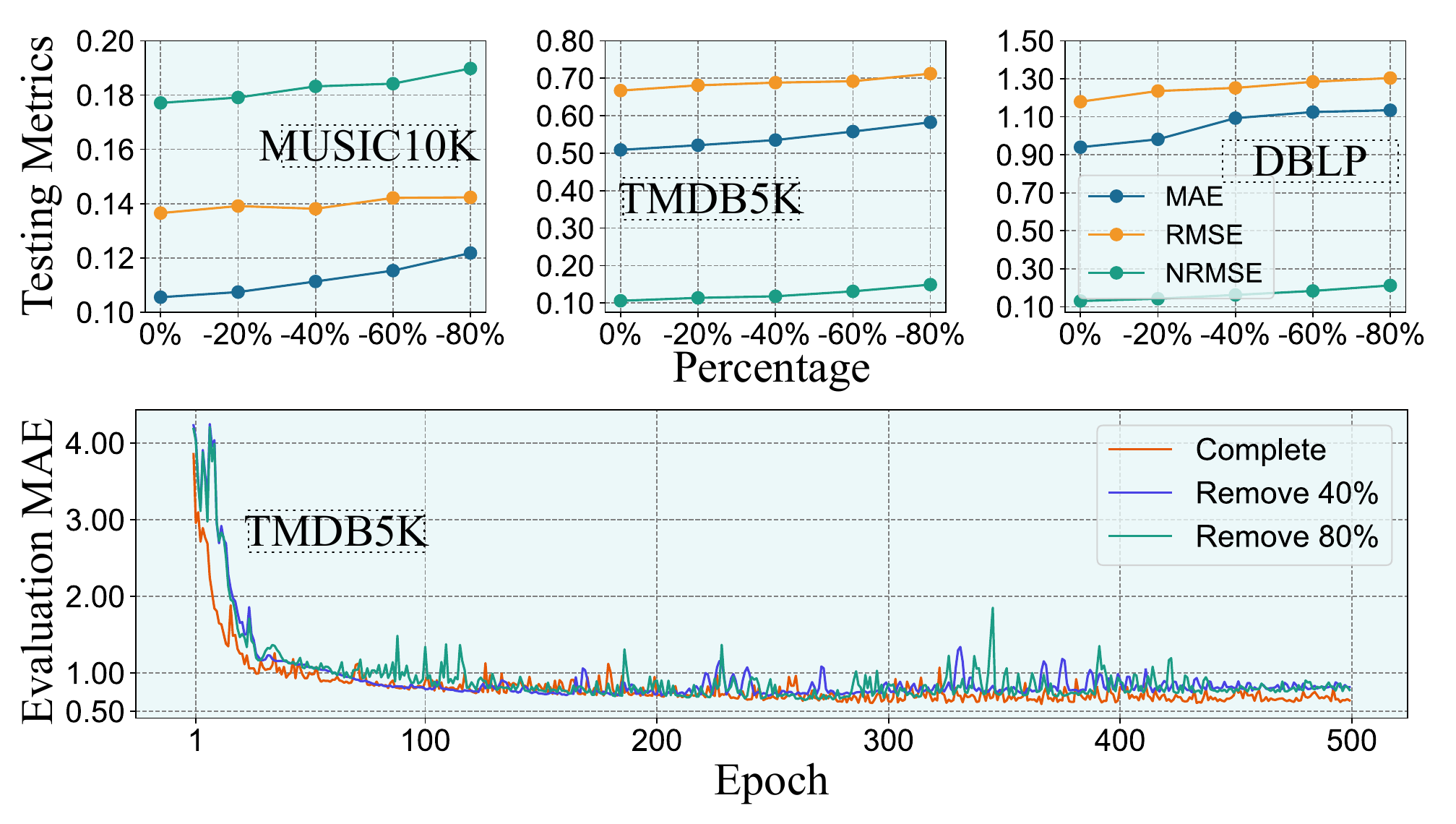}
    \caption{(1) Decreasing contribution of structural knowledge; (2) curves of evaluation MAE (best view in color).}
    \label{fig:vary}
\end{figure}

\begin{table}[t]
\centering
\fontsize{9.5}{9.5} \selectfont
\setlength{\tabcolsep}{1.mm}{
\begin{tabular}
{c|c|c|c|c|c|c}
\specialrule{.1em}{0em}{0em}
\multirow{2}{*}{Variant}           & \multicolumn{3}{c|}{Movie} & \multicolumn{3}{c}{Director} \\ \cline{2-7}
 & MSE & RMSE & NRMSE & MSE & RMSE & NRMSE \\\cline{1-7}
\texttt{w/o} \texttt{WD}    &{0.540}&{0.692}&{0.120}&{0.284}&{0.523}&{0.092} \\ \cline{1-7}
\texttt{w/o} \texttt{$\lambda$}    &{0.524}&{0.675}&{0.116}&{0.279}&{0.519}&{0.092} \\ \cline{1-7}
\model                      &\textbf{0.509}&\textbf{0.667}&\textbf{0.106}&\textbf{0.274}&\textbf{0.507}&\textbf{0.083} \\
\specialrule{.1em}{0em}{0em}
\end{tabular}}
\caption{Study of node importance estimation.}
\label{tab:importance_study}
\end{table}

\begin{figure}[t]
    \centering
    \includegraphics[width=1\linewidth]{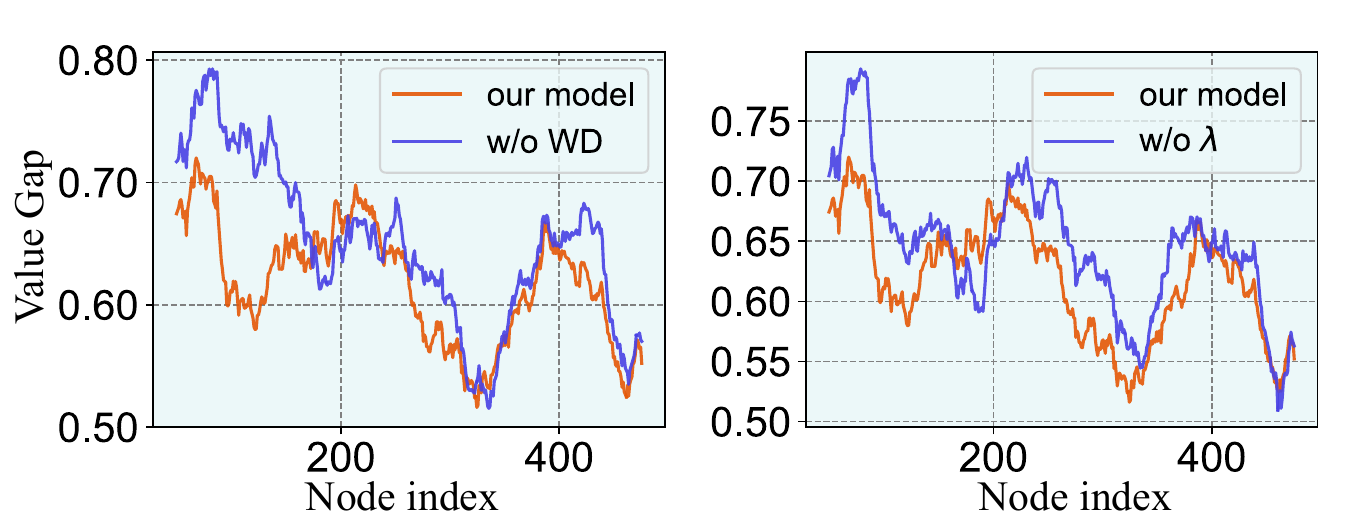}
    \caption{Absolute importance value gap (best view in color)} 
    \label{fig:gap}
\end{figure}

\subsection{Study of Knowledge Synergy and Measurement (RQ2)}
\textbf{\emph{1) Synergy of Structural Knowledge.}}
To validate the contribution of structural knowledge, we randomly disable the knowledge proportion, i.e., {\small $\mathbf c_{i,k}^{(l)}$}, from 0\% (intact) to -80\%.
Our observations from Figure~\ref{fig:vary} are twofold:
(1) From the upper-row figures, we notice remarkable performance degradations across three datasets, where DBLP presents a more pronounced performance perturbation on MAE and RMSE curves.
This demonstrates the efficacy of our knowledge synergy mechanism in identifying node importance, especially in larger HINs with diverse and complex structural information.
(2) We plot MAE curves for cases of \textit{keeping intact}, \textit{removing 40\%}, and \textit{removing 80\%} prior knowledge in the first 500 training epochs on ``\textit{Movie}'' of TMDB5K.
As shown in the lower-row figure, we observe that \model~produces much less bursting perturbations than the curves of \textit{removing 40\%} and \textit{removing 80\%}.
This implies that our implementation to adaptively fuse heterogeneous knowledge can also help to stabilize the model performance.


\textbf{\emph{2) Mechanism of Node Importance Estimation.}}
To evaluate the effectiveness of our proposed importance estimation method, we design two variants on TMDB5K.
(1) Firstly, we replace our original design with 1-Wasserstein distance by simply utilizing a two-layer of MLP for node importance estimation. 
We denote the variant as \texttt{w/o} \texttt{WD}.
(2) Secondly, we retain the measurement of 1-Wasserstein distance but remove $\emb{\lambda}$ in Eqn.~(\ref{eq:score}) for regression, denoted as \texttt{w/o} \texttt{$\lambda$}.
Results in Table~\ref{tab:importance_study} not only justify that integrating Optimal Transport theory for importance estimation achieves better performance (comparing \texttt{w/o} \texttt{WD} to \texttt{w/o} \texttt{$\lambda$}), but also prove the simplicity yet effectiveness of non-linear value regression with learnable $\emb{\lambda}$ (comparing \model~to \texttt{w/o} \texttt{$\lambda$}).
We further visualize the absolute value gap between the ground truth and estimated values of ``\textit{Movie}'' nodes.
To provide a readable visualization in Figure~\ref{fig:gap}, the curves are based on averaged values with the rolling window as 10 nodes.
As the curve gets closer to the bottom, the value estimation will be more accurate.
The curves in Figure~\ref{fig:gap} indicate that \model~consistently performs better than those two variants.


%% file: sections/exp/ablation.tex
\input{sections/tables/ablation}

\subsection{Ablation Study (RQ3)}

\textbf{\emph{1) Network Heterogeneity Learning.}}
We adopt metapath-based methodology to learn heterogeneity.
To validate its usefulness, we propose a variant \texttt{w/o} \texttt{NH} via directly treating input networks as homogeneous.
From Table~\ref{tab:ablation}, we observe over 8.33\% (MSE of ``\textit{Movie}'' type) performance decay on TMDB5K.
This demonstrates the necessity of explicitly distinguishing the heterogeneity of node and edge types for the node importance estimation task.

\textbf{\emph{2) Self-attention in Synergetic Representation Learning.}}
We create a variant \texttt{w/o} \texttt{ATT} by disabling all designs in Eqn's.~(\ref{eq:trsfm}-\ref{eq:trsfm_iter}) and simply using $\mathbf e_{i}$ to replace $\mathbf h^R_{i}$ for importance estimation.
The empirical results between \texttt{w/o} \texttt{ATT} and \model~in Table~\ref{tab:ablation} clearly demonstrate that the self-attention mechanism also work for our model to attentively adjust the contributions for different sources of structural prior knowledge in model learning.

%% file: sections/tables/ablation.tex
\begin{table}[t]
\centering
\fontsize{9.5}{9.5} \selectfont
\setlength{\tabcolsep}{1.mm}{
\begin{tabular}
{c|c|c|c|c|c|c}
\specialrule{.1em}{0em}{0em}
\multirow{2}{*}{Variant}           & \multicolumn{3}{c|}{Movie} & \multicolumn{3}{c}{Director} \\ \cline{2-7}
 & MSE & RMSE & NRMSE & MSE & RMSE & NRMSE \\\cline{1-7}
\texttt{w/o} \texttt{NH} 	&{0.551}&{0.722}&{0.113}&{0.320}&{0.573}&{0.098} \\ \cline{1-7}
\texttt{w/o} \texttt{ATT} 	&{0.544}&{0.696}&{0.118}&{0.319}&{0.549}&{0.096} \\ \cline{1-7}
\model 						&\textbf{0.509}&\textbf{0.667}&\textbf{0.106}&\textbf{0.274}&\textbf{0.507}&\textbf{0.083} \\
\specialrule{.1em}{0em}{0em}
\end{tabular}
}
\caption{Ablation study of other module designs.}
\label{tab:ablation}
\end{table}

%% file: sections/con.tex
\section{Conclusion and Future Work}
We propose a novel framework \model~for estimating HIN node importance.
\model~effectively leverages structural knowledge to harness information synergy, providing a robust measurement of node importance.
Our empirical model evaluation on three public benchmarks demonstrates the performance superiority of \model~against competing baselines.
As for future work, we identify two promising directions:
(1) It is worth investigating other Learning paradigm~\cite{zhang2022costa,zhang2023spectral,zhang2023contrastive,song2023optimal,song2023no,he2023offline} to further improve the quality of learned node embeddings from heterogeneous information. 
(2) We also plan to incorporate \textit{language/vision} modeling~\cite{QiuSYS22,QiuSOYC21,chen2022effective,li2022text,sun2022unified,li2020unsupervised,hu2021semi,hu2023large,zhu2023pointclip} as practical HINs may contain multi-modal information.

%% file: main.bbl
\begin{thebibliography}{65}
\providecommand{\natexlab}[1]{#1}

\bibitem[{Chen et~al.(2023{\natexlab{a}})Chen, Fang, Zhang, and King}]{bgch}
Chen, Y.; Fang, Y.; Zhang, Y.; and King, I. 2023{\natexlab{a}}.
\newblock Bipartite Graph Convolutional Hashing for Effective and Efficient
  Top-N Search in Hamming Space.
\newblock In \emph{WWW}, 3164--3172.

\bibitem[{Chen et~al.(2022{\natexlab{a}})Chen, Guo, Zhang, Ma, Tang, Li, and
  King}]{bgr}
Chen, Y.; Guo, H.; Zhang, Y.; Ma, C.; Tang, R.; Li, J.; and King, I.
  2022{\natexlab{a}}.
\newblock Learning binarized graph representations with multi-faceted
  quantization reinforcement for top-k recommendation.
\newblock In \emph{SIGKDD}, 168--178.

\bibitem[{Chen et~al.(2023{\natexlab{b}})Chen, Truong, Shen, Wang, Li, Chan,
  and King}]{chen2023topological}
Chen, Y.; Truong, T.; Shen, X.; Wang, M.; Li, J.; Chan, J.; and King, I.
  2023{\natexlab{b}}.
\newblock Topological representation learning for e-commerce shopping
  behaviors.
\newblock In \emph{MLG-KDD}.

\bibitem[{Chen et~al.(2020)Chen, Wu, Ma, and King}]{chen2020literature}
Chen, Y.; Wu, Y.; Ma, S.; and King, I. 2020.
\newblock A Literature Review of Recent Graph Embedding Techniques for
  Biomedical Data.
\newblock In \emph{ICONIP}, 21--29.

\bibitem[{Chen et~al.(2022{\natexlab{b}})Chen, Yang, Zhang, Zhao, Meng, Hao,
  and King}]{lkgr}
Chen, Y.; Yang, M.; Zhang, Y.; Zhao, M.; Meng, Z.; Hao, J.; and King, I.
  2022{\natexlab{b}}.
\newblock Modeling scale-free graphs with hyperbolic geometry for
  knowledge-aware recommendation.
\newblock In \emph{WSDM}, 94--102.

\bibitem[{Chen et~al.(2022{\natexlab{c}})Chen, Yang, Wang, Bai, Song, and
  King}]{chen2022attentive}
Chen, Y.; Yang, Y.; Wang, Y.; Bai, J.; Song, X.; and King, I.
  2022{\natexlab{c}}.
\newblock Attentive knowledge-aware graph convolutional networks with
  collaborative guidance for personalized recommendation.
\newblock In \emph{ICDE}, 299--311. IEEE.

\bibitem[{Chen et~al.(2021)Chen, Zhang, Fang, Cao, and
  King}]{chen2021efficient}
Chen, Y.; Zhang, J.; Fang, Y.; Cao, X.; and King, I. 2021.
\newblock Efficient community search over large directed graphs: An augmented
  index-based approach.
\newblock In \emph{IJCAI}, 3544--3550.

\bibitem[{Chen et~al.(2022{\natexlab{d}})Chen, Zhang, Guo, Tang, and
  King}]{chen2022effective}
Chen, Y.; Zhang, Y.; Guo, H.; Tang, R.; and King, I. 2022{\natexlab{d}}.
\newblock An Effective Post-training Embedding Binarization Approach for Fast
  Online Top-K Passage Matching.
\newblock In \emph{AACL}, 102--108.

\bibitem[{Chen et~al.(2023{\natexlab{c}})Chen, Zhang, Yang, Song, Ma, and
  King}]{chen2023wsfe}
Chen, Y.; Zhang, Y.; Yang, M.; Song, Z.; Ma, C.; and King, I.
  2023{\natexlab{c}}.
\newblock WSFE: Wasserstein Sub-graph Feature Encoder for Effective User
  Segmentation in Collaborative Filtering.
\newblock In \emph{SIGIR}, 2521--2525.

\bibitem[{Dorogovtsev, Goltsev, and Mendes(2006)}]{dorogovtsev2006k}
Dorogovtsev, S.~N.; Goltsev, A.~V.; and Mendes, J. F.~F. 2006.
\newblock K-core organization of complex networks.
\newblock \emph{Physical review letters}, 96(4): 040601.

\bibitem[{Egghe et~al.(2006)}]{egghe2006improvement}
Egghe, L.; et~al. 2006.
\newblock An improvement of the h-index: The g-index.
\newblock \emph{ISSI newsletter}, 2(1): 8--9.

\bibitem[{Fang et~al.(2017)Fang, Cheng, Chen, Luo, and Hu}]{fang2017effective}
Fang, Y.; Cheng, R.; Chen, Y.; Luo, S.; and Hu, J. 2017.
\newblock Effective and efficient attributed community search.
\newblock \emph{The VLDB Journal}, 26: 803--828.

\bibitem[{Fu and King(2023)}]{fu2023fedhgn}
Fu, X.; and King, I. 2023.
\newblock FedHGN: {A} Federated Framework for Heterogeneous Graph Neural
  Networks.
\newblock In \emph{{IJCAI}}, 3705--3713.

\bibitem[{Fu and King(2024)}]{fu2024mecch}
Fu, X.; and King, I. 2024.
\newblock {MECCH:} Metapath Context Convolution-based Heterogeneous Graph
  Neural Networks.
\newblock \emph{Neural Networks}, 170: 266--275.

\bibitem[{Fu et~al.(2020)Fu, Zhang, Meng, and King}]{fu2020magnn}
Fu, X.; Zhang, J.; Meng, Z.; and King, I. 2020.
\newblock Magnn: Metapath aggregated graph neural network for heterogeneous
  graph embedding.
\newblock In \emph{WWW}, 2331--2341.

\bibitem[{Grover and Leskovec(2016)}]{grover2016node2vec}
Grover, A.; and Leskovec, J. 2016.
\newblock node2vec: Scalable feature learning for networks.
\newblock In \emph{SIGKDD}, 855--864.

\bibitem[{Haveliwala(2003)}]{haveliwala2003topic}
Haveliwala, T.~H. 2003.
\newblock Topic-sensitive pagerank: A context-sensitive ranking algorithm for
  web search.
\newblock \emph{TKDE}, 15(4): 784--796.

\bibitem[{He et~al.(2023{\natexlab{a}})He, He, Zhang, Zhang, Tang, and
  Ma}]{he2023dynamic}
He, B.; He, X.; Zhang, R.; Zhang, Y.; Tang, R.; and Ma, C. 2023{\natexlab{a}}.
\newblock Dynamic Embedding Size Search with Minimum Regret for Streaming
  Recommender System.
\newblock In \emph{CIKM}, 741--750.

\bibitem[{He et~al.(2023{\natexlab{b}})He, He, Zhang, Tang, and
  Ma}]{he2023dynamically}
He, B.; He, X.; Zhang, Y.; Tang, R.; and Ma, C. 2023{\natexlab{b}}.
\newblock Dynamically Expandable Graph Convolution for Streaming
  Recommendation.
\newblock In \emph{WWW}, 1457--1467.

\bibitem[{He et~al.(2023{\natexlab{c}})He, Sun, Liu, Zhang, Chen, and
  Ma}]{he2023offline}
He, B.; Sun, Z.; Liu, J.; Zhang, S.; Chen, X.; and Ma, C. 2023{\natexlab{c}}.
\newblock Offline imitation learning with variational counterfactual reasoning.
\newblock \emph{arXiv preprint arXiv:2310.04706}.

\bibitem[{Hirsch(2005)}]{hirsch2005index}
Hirsch, J.~E. 2005.
\newblock An index to quantify an individual's scientific research output.
\newblock \emph{PNAS}, 102(46): 16569--16572.

\bibitem[{Hu et~al.(2023)Hu, Chen, Li, Guo, Wen, Yu, and Guo}]{hu2023large}
Hu, X.; Chen, J.; Li, X.; Guo, Y.; Wen, L.; Yu, P.~S.; and Guo, Z. 2023.
\newblock Do Large Language Models Know about Facts?
\newblock \emph{arXiv preprint arXiv:2310.05177}.

\bibitem[{Hu et~al.(2022)Hu, Guo, Wu, Liu, Wen, and Yu}]{hu2022chef}
Hu, X.; Guo, Z.; Wu, G.; Liu, A.; Wen, L.; and Yu, P.~S. 2022.
\newblock CHEF: A Pilot Chinese Dataset for Evidence-Based Fact-Checking.
\newblock In \emph{NAACL-HLT}, 3362--3376.

\bibitem[{Hu et~al.(2020{\natexlab{a}})Hu, Wen, Xu, Zhang, and
  Yu}]{hu2020selfore}
Hu, X.; Wen, L.; Xu, Y.; Zhang, C.; and Yu, P.~S. 2020{\natexlab{a}}.
\newblock SelfORE: Self-supervised Relational Feature Learning for Open
  Relation Extraction.
\newblock In \emph{EMNLP}, 3673--3682.

\bibitem[{Hu et~al.(2021{\natexlab{a}})Hu, Zhang, Ma, Liu, Wen, and
  Yu}]{hu2021semi}
Hu, X.; Zhang, C.; Ma, F.; Liu, C.; Wen, L.; and Yu, P.~S. 2021{\natexlab{a}}.
\newblock Semi-supervised Relation Extraction via Incremental Meta
  Self-Training.
\newblock In \emph{EMNLP}, 487--496.

\bibitem[{Hu et~al.(2021{\natexlab{b}})Hu, Zhang, Yang, Li, Lin, Wen, and
  Yu}]{hu2021gradient}
Hu, X.; Zhang, C.; Yang, Y.; Li, X.; Lin, L.; Wen, L.; and Yu, P.~S.
  2021{\natexlab{b}}.
\newblock Gradient Imitation Reinforcement Learning for Low Resource Relation
  Extraction.
\newblock In \emph{EMNLP}, 2737--2746.

\bibitem[{Hu et~al.(2020{\natexlab{b}})Hu, Dong, Wang, and
  Sun}]{hu2020heterogeneous}
Hu, Z.; Dong, Y.; Wang, K.; and Sun, Y. 2020{\natexlab{b}}.
\newblock Heterogeneous graph transformer.
\newblock In \emph{WWW}, 2704--2710.

\bibitem[{Huang et~al.(2022)Huang, Fang, Lin, Cao, Zhang, and
  Orlowska}]{ch2022hiven}
Huang, C.; Fang, Y.; Lin, X.; Cao, X.; Zhang, W.; and Orlowska, M. 2022.
\newblock Estimating Node Importance Values in Heterogeneous Information
  Networks.
\newblock In \emph{ICDE}, 846--858.

\bibitem[{Huang et~al.(2021)Huang, Sun, Du, Liu, Lv, and
  Xiong}]{huang2021representation}
Huang, H.; Sun, L.; Du, B.; Liu, C.; Lv, W.; and Xiong, H. 2021.
\newblock Representation Learning on Knowledge Graphs for Node Importance
  Estimation.
\newblock In \emph{SIGKDD}, 646--655.

\bibitem[{Korotin, Selikhanovych, and
  Burnaev(2023)}]{DBLP:conf/iclr/KorotinSB23}
Korotin, A.; Selikhanovych, D.; and Burnaev, E. 2023.
\newblock Neural Optimal Transport.
\newblock In \emph{ICLR}. OpenReview.net.

\bibitem[{Li et~al.(2022)Li, Li, Ge, King, and Lyu}]{li2022text}
Li, J.; Li, Z.; Ge, T.; King, I.; and Lyu, M.~R. 2022.
\newblock Text Revision by On-the-Fly Representation Optimization.
\newblock In \emph{AAAI}, 10956--10964.

\bibitem[{Li et~al.(2020)Li, Li, Mou, Jiang, Lyu, and
  King}]{li2020unsupervised}
Li, J.; Li, Z.; Mou, L.; Jiang, X.; Lyu, M.; and King, I. 2020.
\newblock Unsupervised text generation by learning from search.
\newblock In \emph{NeurIPS}, volume~33, 10820--10831.

\bibitem[{Liang et~al.(2023)Liang, Hu, Xu, Song, and
  King}]{liang2023predicting}
Liang, L.; Hu, X.; Xu, Z.; Song, Z.; and King, I. 2023.
\newblock Predicting Global Label Relationship Matrix for Graph Neural Networks
  under Heterophily.
\newblock In \emph{NeurIPS}.

\bibitem[{Marchiori and Latora(2000)}]{marchiori2000harmony}
Marchiori, M.; and Latora, V. 2000.
\newblock Harmony in the small-world.
\newblock \emph{Physica A: Statistical Mechanics and its Applications},
  285(3-4): 539--546.

\bibitem[{Naderializadeh et~al.(2021)Naderializadeh, Comer, Andrews, Hoffmann,
  and Kolouri}]{pswe}
Naderializadeh, N.; Comer, J.~F.; Andrews, R.; Hoffmann, H.; and Kolouri, S.
  2021.
\newblock Pooling by sliced-Wasserstein embedding.
\newblock volume~34, 3389--3400.

\bibitem[{Negre et~al.(2018)Negre, Morzan, Hendrickson, Pal, Lisi, Loria,
  Rivalta, Ho, and Batista}]{negre2018eigenvector}
Negre, C.~F.; Morzan, U.~N.; Hendrickson, H.~P.; Pal, R.; Lisi, G.~P.; Loria,
  J.~P.; Rivalta, I.; Ho, J.; and Batista, V.~S. 2018.
\newblock Eigenvector centrality for characterization of protein allosteric
  pathways.
\newblock \emph{PNAS}, 115(52): E12201--E12208.

\bibitem[{Nieminen(1974)}]{nieminen1974centrality}
Nieminen, J. 1974.
\newblock On the centrality in a graph.
\newblock \emph{Scandinavian journal of psychology}, 15(1): 332--336.

\bibitem[{Nietert et~al.(2022)Nietert, Goldfeld, Sadhu, and
  Kato}]{nietert2022statistical}
Nietert, S.; Goldfeld, Z.; Sadhu, R.; and Kato, K. 2022.
\newblock Statistical, robustness, and computational guarantees for sliced
  wasserstein distances.
\newblock \emph{NeurIPS}, 35: 28179--28193.

\bibitem[{Page et~al.(1999)Page, Brin, Motwani, and
  Winograd}]{page1999pagerank}
Page, L.; Brin, S.; Motwani, R.; and Winograd, T. 1999.
\newblock The PageRank citation ranking: Bringing order to the web.
\newblock Technical report, Stanford InfoLab.

\bibitem[{Park et~al.(2019)Park, Kan, Dong, Zhao, and
  Faloutsos}]{park2019estimating}
Park, N.; Kan, A.; Dong, X.~L.; Zhao, T.; and Faloutsos, C. 2019.
\newblock Estimating node importance in knowledge graphs using graph neural
  networks.
\newblock In \emph{SIGKDD}, 596--606.

\bibitem[{Park et~al.(2020)Park, Kan, Dong, Zhao, and
  Faloutsos}]{park2020multiimport}
Park, N.; Kan, A.; Dong, X.~L.; Zhao, T.; and Faloutsos, C. 2020.
\newblock Multiimport: Inferring node importance in a knowledge graph from
  multiple input signals.
\newblock In \emph{SIGKDD}, 503--512.

\bibitem[{Qiu et~al.(2021)Qiu, Su, Ou, Yu, and Chen}]{QiuSOYC21}
Qiu, Z.; Su, Q.; Ou, Z.; Yu, J.; and Chen, C. 2021.
\newblock Unsupervised Hashing with Contrastive Information Bottleneck.
\newblock In \emph{{IJCAI}}, 959--965.

\bibitem[{Qiu et~al.(2022)Qiu, Su, Yu, and Si}]{QiuSYS22}
Qiu, Z.; Su, Q.; Yu, J.; and Si, S. 2022.
\newblock Efficient Document Retrieval by End-to-End Refining and Quantizing
  {BERT} Embedding with Contrastive Product Quantization.
\newblock In \emph{{EMNLP}}, 853--863.

\bibitem[{Reimers and Gurevych(2019)}]{reimers2019sentence}
Reimers, N.; and Gurevych, I. 2019.
\newblock Sentence-bert: Sentence embeddings using siamese bert-networks.
\newblock \emph{arXiv preprint arXiv:1908.10084}.

\bibitem[{Sabidussi(1966)}]{sabidussi1966centrality}
Sabidussi, G. 1966.
\newblock The centrality index of a graph.
\newblock \emph{Psychometrika}, 31(4): 581--603.

\bibitem[{Shaw(1954)}]{shaw1954group}
Shaw, M.~E. 1954.
\newblock Group structure and the behavior of individuals in small groups.
\newblock \emph{The Journal of psychology}, 38(1): 139--149.

\bibitem[{Song, Zhang, and King(2023{\natexlab{a}})}]{song2023no}
Song, Z.; Zhang, Y.; and King, I. 2023{\natexlab{a}}.
\newblock No Change, No Gain: Empowering Graph Neural Networks with Expected
  Model Change Maximization for Active Learning.
\newblock In \emph{NeurIPS}.

\bibitem[{Song, Zhang, and King(2023{\natexlab{b}})}]{song2023optimal}
Song, Z.; Zhang, Y.; and King, I. 2023{\natexlab{b}}.
\newblock Optimal Block-wise Asymmetric Graph Construction for Graph-based
  Semi-supervised Learning.
\newblock In \emph{NeurIPS}.

\bibitem[{Song, Zhang, and King(2023{\natexlab{c}})}]{DBLP:conf/cikm/SongZK23}
Song, Z.; Zhang, Y.; and King, I. 2023{\natexlab{c}}.
\newblock Towards Fair Financial Services for All: {A} Temporal {GNN} Approach
  for Individual Fairness on Transaction Networks.
\newblock In \emph{{CIKM}}, 2331--2341. {ACM}.

\bibitem[{Sun et~al.(2022)Sun, Ge, Ma, Li, Wei, and Wang}]{sun2022unified}
Sun, X.; Ge, T.; Ma, S.; Li, J.; Wei, F.; and Wang, H. 2022.
\newblock A Unified Strategy for Multilingual Grammatical Error Correction with
  Pre-trained Cross-Lingual Language Model.
\newblock In Raedt, L.~D., ed., \emph{IJCAI}, 4367--4374.
\newblock Main Track.

\bibitem[{Sun et~al.(2011)Sun, Han, Yan, Yu, and Wu}]{sun2011pathsim}
Sun, Y.; Han, J.; Yan, X.; Yu, P.~S.; and Wu, T. 2011.
\newblock Pathsim: Meta path-based top-k similarity search in heterogeneous
  information networks.
\newblock \emph{VLDB}, 4(11): 992--1003.

\bibitem[{Vaswani et~al.(2017)Vaswani, Shazeer, Parmar, Uszkoreit, Jones,
  Gomez, Kaiser, and Polosukhin}]{vaswani2017attention}
Vaswani, A.; Shazeer, N.; Parmar, N.; Uszkoreit, J.; Jones, L.; Gomez, A.~N.;
  Kaiser, {\L}.; and Polosukhin, I. 2017.
\newblock Attention is all you need.
\newblock \emph{NeurIPS}, 30.

\bibitem[{Veli{\v{c}}kovi{\'c} et~al.(2017)Veli{\v{c}}kovi{\'c}, Cucurull,
  Casanova, Romero, Lio, and Bengio}]{velivckovic2017graph}
Veli{\v{c}}kovi{\'c}, P.; Cucurull, G.; Casanova, A.; Romero, A.; Lio, P.; and
  Bengio, Y. 2017.
\newblock Graph attention networks.
\newblock \emph{arXiv preprint arXiv:1710.10903}.

\bibitem[{Villani et~al.(2009)}]{villani2009optimal}
Villani, C.; et~al. 2009.
\newblock \emph{Optimal transport: old and new}, volume 338.
\newblock Springer.

\bibitem[{Yang et~al.(2022)Yang, Li, Zhou, Liu, and King}]{yang2022hicf}
Yang, M.; Li, Z.; Zhou, M.; Liu, J.; and King, I. 2022.
\newblock Hicf: Hyperbolic informative collaborative filtering.
\newblock In \emph{SIGKDD}, 2212--2221.

\bibitem[{Yang et~al.(2023{\natexlab{a}})Yang, Zhou, Pan, and
  King}]{yang2023kappahgcn}
Yang, M.; Zhou, M.; Pan, L.; and King, I. 2023{\natexlab{a}}.
\newblock $\kappa$HGCN: Tree-likeness Modeling via Continuous and Discrete
  Curvature Learning.
\newblock In \emph{SIGKDD}, 2965--2977.

\bibitem[{Yang et~al.(2023{\natexlab{b}})Yang, Zhou, Ying, Chen, and
  King}]{yang2023hyperbolic}
Yang, M.; Zhou, M.; Ying, R.; Chen, Y.; and King, I. 2023{\natexlab{b}}.
\newblock Hyperbolic Representation Learning: Revisiting and Advancing.
\newblock \emph{ICML}.

\bibitem[{Zhang et~al.(2022{\natexlab{a}})Zhang, Chen, Gao, Liao, Zhao, and
  King}]{zhang2022knowledge}
Zhang, X.; Chen, Y.; Gao, C.; Liao, Q.; Zhao, S.; and King, I.
  2022{\natexlab{a}}.
\newblock Knowledge-aware Neural Networks with Personalized Feature Referencing
  for Cold-start Recommendation.
\newblock \emph{arXiv preprint arXiv:2209.13973}.

\bibitem[{Zhang et~al.(2023{\natexlab{a}})Zhang, Chen, Song, and
  King}]{zhang2023contrastive}
Zhang, Y.; Chen, Y.; Song, Z.; and King, I. 2023{\natexlab{a}}.
\newblock Contrastive cross-scale graph knowledge synergy.
\newblock In \emph{SIGKDD}, 3422--3433.

\bibitem[{Zhang and Zhu(2019)}]{zhang2019doc2hash}
Zhang, Y.; and Zhu, H. 2019.
\newblock Doc2hash: Learning discrete latent variables for documents retrieval.
\newblock In \emph{NAACL}.

\bibitem[{Zhang et~al.(2023{\natexlab{b}})Zhang, Zhu, Chen, Song, Koniusz, King
  et~al.}]{zhang2023mitigating}
Zhang, Y.; Zhu, H.; Chen, Y.; Song, Z.; Koniusz, P.; King, I.; et~al.
  2023{\natexlab{b}}.
\newblock Mitigating the Popularity Bias of Graph Collaborative Filtering: A
  Dimensional Collapse Perspective.
\newblock In \emph{NeurIPS}.

\bibitem[{Zhang et~al.(2022{\natexlab{b}})Zhang, Zhu, Song, Koniusz, and
  King}]{zhang2022costa}
Zhang, Y.; Zhu, H.; Song, Z.; Koniusz, P.; and King, I. 2022{\natexlab{b}}.
\newblock COSTA: covariance-preserving feature augmentation for graph
  contrastive learning.
\newblock In \emph{SIGKDD}.

\bibitem[{Zhang et~al.(2023{\natexlab{c}})Zhang, Zhu, Song, Koniusz, and
  King}]{zhang2023spectral}
Zhang, Y.; Zhu, H.; Song, Z.; Koniusz, P.; and King, I. 2023{\natexlab{c}}.
\newblock Spectral feature augmentation for graph contrastive learning and
  beyond.
\newblock In \emph{AAAI}.

\bibitem[{Zheng et~al.(2021)Zheng, Zhang, Chen, Zhang, Yang, and
  Wang}]{zheng2021convolutional}
Zheng, Y.; Zhang, X.; Chen, S.; Zhang, X.; Yang, X.; and Wang, D. 2021.
\newblock When Convolutional Network Meets Temporal Heterogeneous Graphs: An
  Effective Community Detection Method.
\newblock \emph{IEEE TKDE}.

\bibitem[{Zhu et~al.(2023)Zhu, Zhang, He, Guo, Zeng, Qin, Zhang, and
  Gao}]{zhu2023pointclip}
Zhu, X.; Zhang, R.; He, B.; Guo, Z.; Zeng, Z.; Qin, Z.; Zhang, S.; and Gao, P.
  2023.
\newblock Pointclip v2: Prompting clip and gpt for powerful 3d open-world
  learning.
\newblock In \emph{ICCV}, 2639--2650.

\end{thebibliography}
